# Pele's hairs and exotic multiply twinned graphite closed-shell microcrystals in meteoritic dust of Chelyabinsk superbolide

Sergey Taskaev[a,b,c]*, Konstantin Skokov[d], Vladimir Khovaylo[a,b], Wolfgang Donner[d], Tom Faske[d], Alexander Dudorov[a], Nick Gorkavyi[e,f], Galina Savosteenko[a], Alexander Dyakonov[c], Woohyeon Baek[g], Artem Kuklin[g], Pavel Avramov[g] and Oliver Gutfleisch[d]

[a] *Chelyabinsk State University (Chelyabinsk, Russia)*
[b] *National University of Science and Technology "MISiS" (Moscow, Russia)*
[c] *NRU South Ural State University (Chelyabinsk, Russia)*
[d] *TU Darmstadt (Darmstadt, Germany)*
[e] *SSAI/GSFC/NASA (MD, USA)*
[f] *Crimean Astrophysical Observatory (Crimea, Russia)*
[g] *Kyungpook National University (Daegu, South Korea)*

When a space body enters Earth's atmosphere, its surface is exposed to high pressure and temperatures. The airflow tears off small droplets from the meteoroid forming a cloud of meteorite dust. Can new materials be synthesized in these unique conditions (high temperature, pressure, gaseous atmosphere, catalysts)? As a rule, meteoritic dust dissipates in the atmosphere without a trace or is mixed with terrestrial soil. The Chelyabinsk superbolide, the biggest in the 21$^{st}$ century, which exploded on February 15, 2013 above snowy fields of the Southern Urals, was an exception. Two new types of materials were found during an in-depth study: thread-like structures that were not previously associated with meteorite falls and unique carbon crystals with a size of several micrometers that were not observed before. The nature of formation of thread-like structures is fully similar to the nature of formation of Pele's hair during eruptions of terrestrial volcanos. Multiple twin growth mechanism of formation of closed shell graphite microcrystals was proposed based on DFT and classical/ab initio MD simulations. It was found that among several possible embryo carbon nanoclusters the $C_{60}$ fullerene and polyhexacyclooctadecane -$C_{18}H_{12}$- may be the main suspects responsible for the formation of closed shell quasi-spherical and hexagonal rod graphite microcrystals.

The superbolide that fell on February 15, 2013 in the area of Chelyabinsk (Russia) was a unique phenomenon in terms of its scale and caused an immense public and scientific interest. It has been the biggest meteoroid in the 21$^{st}$ century to date and the biggest bolide after the Tunguska event. On the one hand, the fall of that space body, which had an initial diameter of about 18 m, showed the absolute lack of defense of our planet from the meteorite hazard and, on the other hand, it brought to our planet unique materials synthesized in the conditions that cannot be reproduced in the advanced laboratories (enormous temperatures, pressures, streams of elementary particles, microgravitation, long isothermal transitions, magnetic fields, catalysts and etc.). The energy release during the fall of that superbolide was equivalent to about 500 kt of TNT, which is approximately 25 larger than the energy of the atomic bombs exploded in Hiroshima and Nagasaki in 1945. Fortunately, as this energy was released along the falling trajectory about 280 km long, there were no casualties or significant damage.

Several dozen thousand tons of space matter falls to Earth annually: these are micrometeorites, which are fully or partially remelted, and space dust with diameters of particles less than 50 micrometers that freely falls not affected by high temperatures [1-3]. These materials present a special interest because their matter could come to Earth both from near space (as is the case with meteorite falls) and from outer space (as is the case with material from gas-dust tails of comets). However, despite the "large" quantity of falling matter, there is a problem with the collection of dust samples: the specific concentration of such materials is very low. The following acclaimed successful studies of such materials can be highlighted: several magnetic spheres were found in marine deposits more than a century ago by the H.M.S. Challenger expedition [4]. According to the elements contained, they were classed as fragments of chondrites. It is difficult to determine the extra-terrestrial origin of microspheres, whose phase composition includes olivine, glasses, magnetite, etc.; such determination is based on the comparison of proportions of elements with the known meteorites [5, 6]. Microspheres of space origin were found in Siberian swamps [7], deserts [8], beach sand [9], deep-sea sediments [10], sedimentary rocks [11-13] in Greenland's cryoconites and thaw ice [14], Antarctic sediments [15], and in glacier ice [16].

The fall of the Chelyabinsk meteorite, which is an ordinary chondrite of the LL5 petrological class, was accompanied by its significant destruction resulting in the falling to Earth's surface of a large number of fragments. The meteorite's disintegration was accompanied by the formation of a gas-dust plume and subsequent settlement of the dust component. According to assessments made before, about 99% of the weight of the meteoroid went to the gas-dust component, whilst the remaining part fell as meteorites.

According to O.P. Popova et al. [17], the dust plume formed at altitudes 80 km to 27 km. It was also detected by NASA's Suomi satellite at altitudes 30-67 km (see Fig. 1), and by other satellites (Japanese satellite MTSAT-2, Russian satellite Elektron, European satellites METEOSAT-9 and METEOSAT-10, Chinese satellite FY-2D, and others). It moved eastward during its evolution and circumnavigated the entire globe in four days. The dust cloud could have been observed for at least three months and subsequently it dissipated [18].

The conditions in which the meteoritic dust fell out could be viewed as unique: there had been a snowfall 8 days before the Chelyabinsk meteorite that created a distinct borderline allowing determination of the layer's beginning; 13 days after the meteorite's fall there also was a snowfall that conserved the meteorite dust that had fallen out by that time (see Fig. 2).

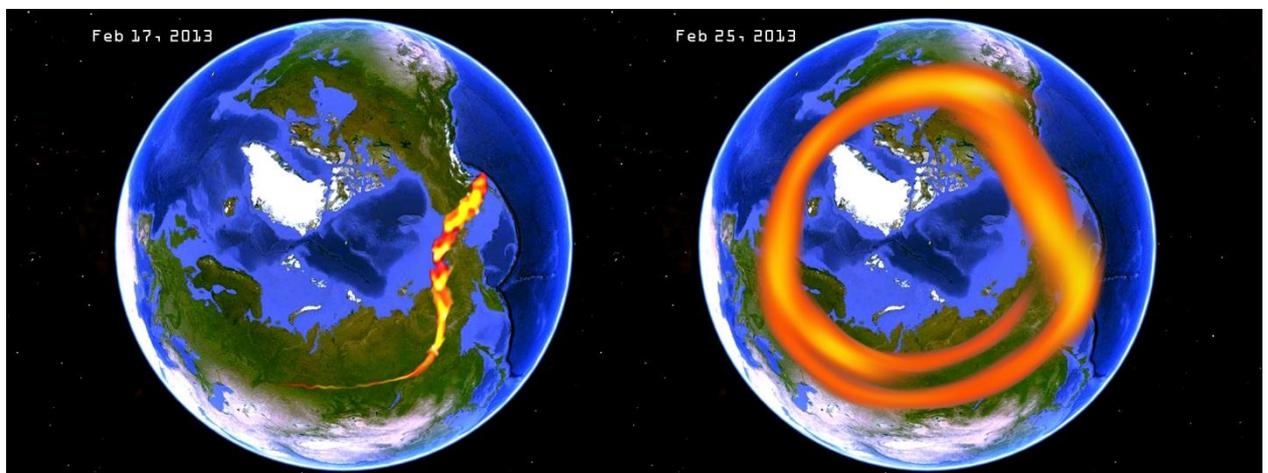

Fig. 1. Model of the dust ring formed after the meteorite's disintegration found by NASA/NOAA's satellite Suomi
(reconstructed with data from [18]). Map data is provided by Google Maps.

This paper is focused on two unique morphological peculiarities of the meteoroid's dust component that have not been observed before: mineral fibers and carbon crystals. Contrary to the expectations that the main dust fraction would be the balls formed during the ablation of matter, the dust samples found in the analyzed samples contained multiple fragments that did not go through thermal differentiation and by their phase composition were close to the chondrite's body; also the samples were found to contain quasi-two-dimensional structures in the form of fibers 0.1-1 mm long and 0.01-0.02 mm wide connecting the drops of the melt (see

Fig. 3 a-d). In terms of mineral composition, these threads were close to pyroxenes with different concentrations of admixtures and consisted of a number of solid solutions $(Na,Ca,Mn)(Mg,Fe,Al)_2[Si_2O_6]$ (see supplementary materials). Formation of mineral threads in nature has been known for a long time. Hawaiian volcanos, which, according to beliefs of the local residents, are governed by the goddess of fire Pele, emit fountains of fiery lava and volcanic bombs. Cold air tears off droplets from the molten rock and draws them into threads with a length of up to two meters forming so-called "Pele's hair". Such thread-like formations are also often found near other volcanos in Africa, Kamchatka, and Iceland. The discovery of mineral threads' formation during meteoroid flights through the atmosphere changes the paradigm of research of meteoritic dust. Mainly spherical particles have been classed as meteoritic dust before, whilst thread-like formations, particles with "tails" and particles connected by "necks" have been classed as man-made or impact formations.

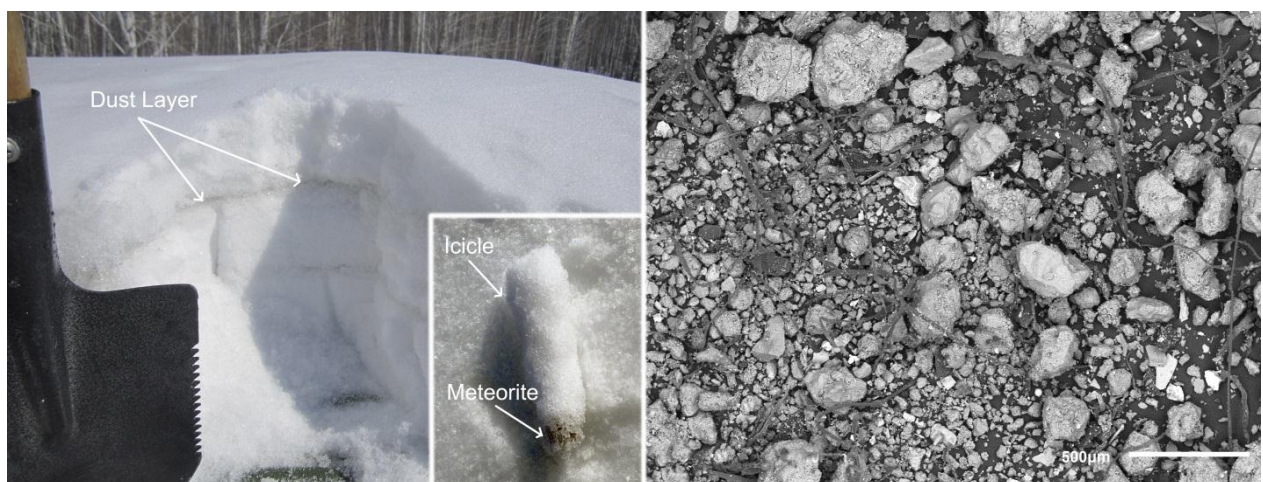

Fig. 2. Layer of meteorite dust in snow (left),
a SEM image of the filtered insoluble fraction (right).

However, Bradley et al. [19] observed microscopic enstatite threads in stratospheric dust samples in 1983, and Noguchi et al. [20] found similar threads in Antarctic snow in 2015. According to authors of these papers [19, 20], such threads formed during the condensation from the gas phase. The origin of loose dust grains and threads collected in snow and in the stratosphere is also attributed by the authors to interplanetary dust or cometary dust. However, our results indicate that such objects can also form during the burning of rather large bolides (several such bodies burn in Earth's atmosphere annually). Their formation is connected with another mechanism: not crystallization from the gas phase but with melting and drawing of the melt, as it happens during the formation of Pele's volcanic hair. It sheds new light on the origin of dust with mineral threads found in the stratosphere, particularly in Antarctica.

Carbon crystals are the second important peculiarity of the dust component found in this study; the first carbon crystal was found during an investigation of the dust using an optical microscope, because its facets happened to be in the focal plane (see Fig. 4 a). Subsequent studies using optical electron microscopy showed that there were a lot of similar objects in the meteoritic dust (see Fig. 4 c, d, and supplementary materials); however, finding them using an electron microscope was rather challenging due to their small size (about 10 μm) and low phase contrast. As seen in Figs. 4, the object has facets with a symmetry axis of the sixth order and reflects light well. Found by the methods of electron microscopy, this object showed a wonderful facet form with quasi-spherical symmetry (see Fig. 4). An EDS analysis showed (see supplementary materials) that the found particle with distinct signs of crystallinity consisted mainly of carbon. This fact explains the difficulties of finding such objects using methods of electron microscopy, because the contrast of particles fully coincide with the contrast created by the electrically conductive polymer base with a high content of carbon used for the installation of samples.

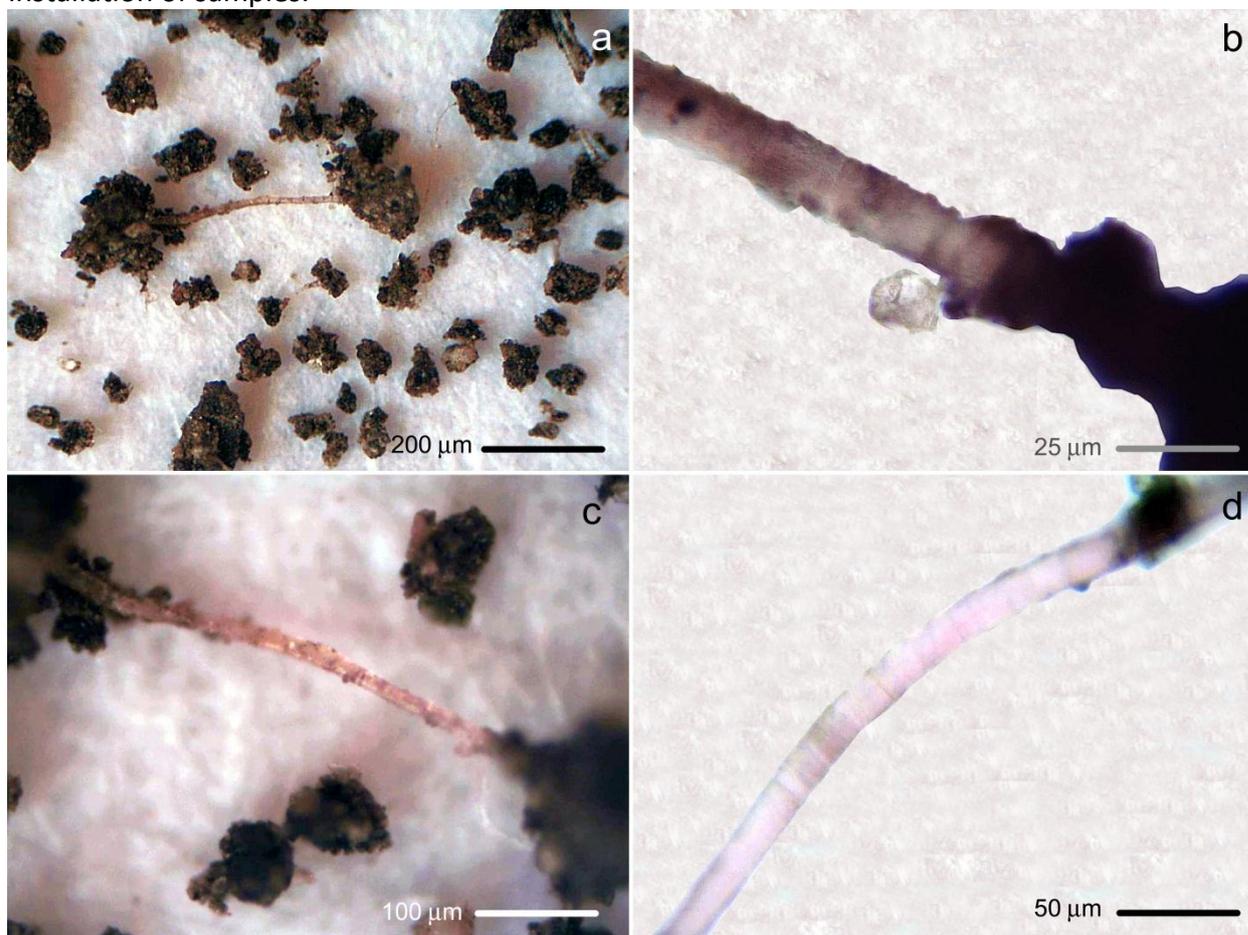

Fig. 3. Mineral fibers in dust of the Chelyabinsk meteorite.

Based on the data available at the time of finding the carbon crystalline particle, the following possible versions of its formation were formulated:

- object is a macroscopic particle with the hexagonal form of the lonsdaleite, a variety of diamond [21];
- object is a molecular crystal, whose lattice is composed of fullerenes (fullerite);
- object is a type of carbon that crystallized based on carbon nanostructure (fullerene or nanotube). This inference is supported by structures in Fig. 4 c where the presence of particles of both quasi-spherical and cylindrical symmetry is clearly seen.

The first version of formation of such particles is supported by the chemical composition, facets, and symmetry axis of the sixth order not observed in crystals of cubic diamonds. However, there is an opinion that the hexagonal crystalline structure of lonsdaleite does not exist in the macroscopic scale and that it is a defect of the packing of the face-centered lattice [22]. Raman-spectroscopy data could confirm the belonging to this exotic structure if such data confirm that carbon atoms in this particle are held in the $sp^3$ hybridized state. However, our study reveals that the peak at 1330 cm$^{-1}$, which is characteristic of the $sp^3$ hybridized carbon, was completely absent in the spectrum of the first particle, whilst peaks fully equivalent to graphite (G and D modes) were present. However, the D-mode peak at the frequency of 1330 cm$^{-1}$ was found in the spectra of other particles (see supplementary materials), which could have been caused by multiple defects of the structure and facets, and it would not be right to say that the particle in this case could have a heterogenic structure composed of phase mixtures represented by the *sp*² and *sp*³ hybridized atoms.

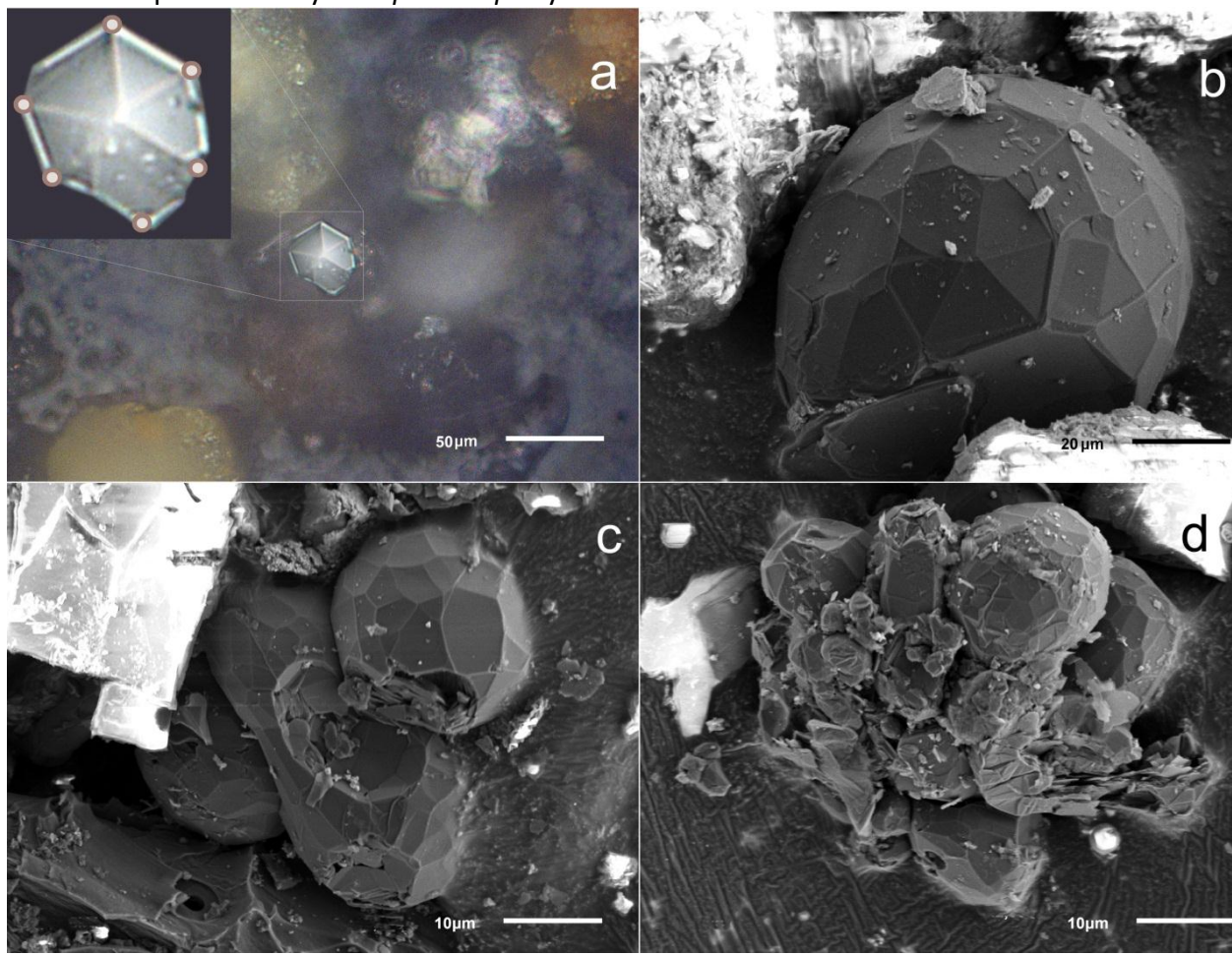

Fig. 4. (a) Optical and (b-d) SEM images of the carbon crystals.

The results of an X-ray structural analysis gave a more accurate answer to this question. In order to perform the X-ray analysis, three particles were taken out of the substrate using a micro-manipulator. Fig. 5 shows that the found carbon particles in the optical range are black and have facets. Today, industrial carbonado diamonds are known—the polycrystalline variety of a cubic diamond with a color varying from black to dark green. Carbonado diamonds are synthesized at a pressure of 8-12 GPa, which is easily reachable in the explosion of a bolide in the atmosphere. Note that the formation of carbonado diamonds is catalyzed by nickel that was also present in the kamacite of the meteorite.

X-ray diffraction patterns were collected for all extracted carbon particles (see Supplementary Materials for detailed information for each particle). It was found that the particles were no single crystals, and that the majority of reflections were observed at interplanar spacings of $d \approx 3.36$ Å. This particular lattice spacing is signature for graphite (002) planes. The spherical arrangement of (002) reflections in reciprocal space resembles the facetted morphology of the carbon particles in real space, see Fig. 5(b). It can be assumed, that the facets correspond to individual (001) planes. The particle, therefore, is comprised of hundreds of facets with the typical bulk structure for graphite ($P6_3mc$, a= 2.46 Å, c = 6.72 Å). For the largest carbon particle, see Supplementary Figure SI5-11, the diffraction pattern has, in addition to the spherically distributed graphite (002) planes, a complete set of reflections of single crystalline graphite. Concluding from this, the larger particles are agglomerates of more loosely bound graphite sheets. The absence of XRD powder rings for all particles indicates that the macroscopic faceting continues down to the atomic scale, which excludes having a carbon onion structure inside. Due to the small size of the analyzed objects, obtaining high-quality diffractograms was very difficult; however, based on the obtained data, we can confidently state that the analyzed particles were exotically-shaped graphite.

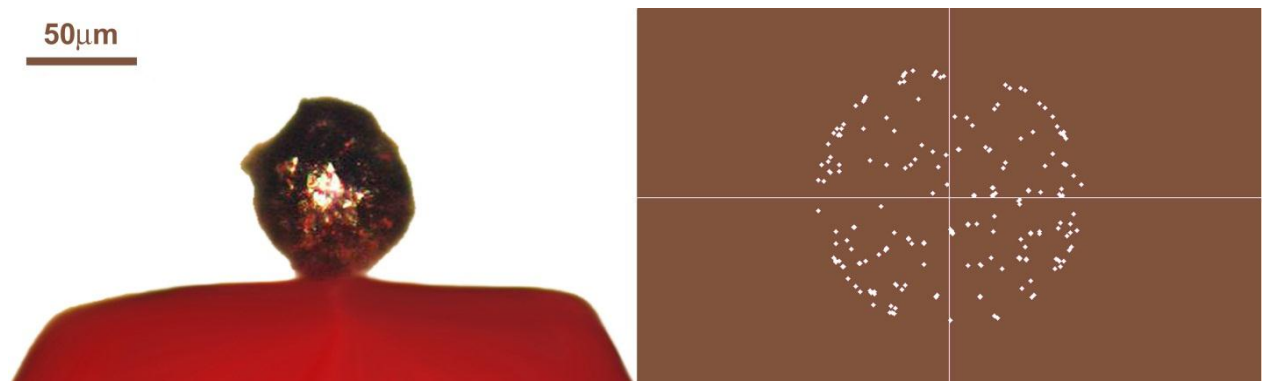

Fig. 5. (left) Optical image of recovered carbon particle glued on epoxy resin.
(right) Macroscopically visible facets correspond to crystallographic (001) planes of graphite Arrangement of x-ray reflections of the same particle in reciprocal space. Spherically distributed (002) reflections of graphite with $d \approx 3.36$ Å make up the majority of the diffraction pattern, resembling the facetted shape of the particle.

Since the Chelyabinsk meteorite was not a carbonaceous chondrite and did not contain carbon in significant concentrations, the issue arises about the nature of origin of such particles. The most probable version, in our opinion, is thermal dissociation of $CO_2$ molecules from Earth's atmosphere under the influence of high temperatures of the bolide and shock waves. However, the $CO_2$ molecule is known to be one of the most stable compounds, and its thermal dissociation requires a temperature of about several thousand degrees. Such temperatures are quite possible during a bolide's flight and the conducted numeric simulation [23] shows that temperatures on the shock front of the bolide rise above $10^4$ K, and a cone forms behind the bolide with rather high temperatures (over 500°C). It should be mentioned that there was a series of explosions during the bolide's flight, and temperature also rose by leaps and pressure rose in local fields at the explosion places. The modeling of the superbolide's destruction process is presented at NASA's web site in [24].

Nonetheless, the temperature of thermal dissociation of $CO_2$ molecules can be significantly decreased in the presence of catalysts, whose role iron and nickel can play. As follows from the previous data, the Chelyabinsk meteorite contained about 5% of an iron-nickel alloy represented by inclusions with a size of several hundred micrometers (see supplementary materials). In the presence of such catalysts, the temperature of thermal dissociation could have been locally decreased to 600-700 K, which significantly increases the zone in which pure carbon could form, which, in turn, could have participated in the creation of such crystals.

However, the presence of carbon is a necessary but not sufficient condition for the synthesis of such particles. If the crystallization seed is absent, free carbon must transform into disordered graphite or other allotropic forms of carbon. According to laws of physics, crystal facet formation occurs along the basal planes of the crystal seed represented by the same material from which a crystal is grown. In this connection, it may be assumed that such allotropic forms of carbon as fullerenes or nanotubes could have become the crystallization seed of such structures. As shown previously, both fullerenes and nanotubes in the concentration of several dozen to several hundred ng/m$^3$ occur in the atmosphere even at rather high altitudes [25] and they could have participated in the process of deposition of free carbon to their surface with the formation of facet layers. In particular, fullerene molecules were considered as seed precursors in the synthesis of superhard nanograined, nanotwinned diamond microcrystals with unique mechanical properties [26].

(111) diamond surfaces are the direct precursors for formation of graphene layers during graphitization of diamond crystals [27]. Multiply twinned particles, such as decahedrons and icosahedrons with developed (111) surfaces, are commonly observed morphologies for diamond crystals grown from vapor phase [28-32]. Carbon and silicon fullerenes ($C_{20}/C_{60}$ and $Si_{20}/Si_{60}$), their hydrogenated derivatives ($C_{20}H_{20}$, $C_{60}H_n$ and their silicon analogs), 1D polyhexacyclopentadecane ($-C_{15}H_{10}-$) and polyhexacyclooctadecane ($-C_{18}H_{12}-$) were considered in numerous publications [33-40] as formation embryos of icosahedral and dodecahedral star-shaped nano- and microparticles. Detailed description of structure and properties of nanotwinned nano- and microdiamonds with developed (111) surfaces can be found in the Supplementary Materials. Based on the structural analysis, one can expect that graphitization of multiply twinned nano- and microdiamonds may result in formation of unique closed-shell multiply-twinned quasi-spherical and elongated hexagonal graphite microcrystals presented in Figures 4 (a-d) and 5.

Atomic and electronic structures of icosahedral Class I ($C_{20}$ core-based, Figure SI7-2) and Class II ($C_{60}$ core-based, Figure SI7-4) Goldberg-type nanodiamonds, Class I (Figure SI7-5) and Class II (Figure SI7-6) fullerenes and multiply twinned closed-shell graphite structures (Figures SI7-9, SI7-10), diamond star-shaped dodecahedron nanorods (Figure SI7-7) and hybrid $sp^2/sp^3$ carbon honeycombs (Figures SI7-8, SI7-11) were simulated using Density Functional (DFT), classical Brenner potential and ab initio and classical Molecular Dynamics (MD) techniques. The schematics of graphitization of multiply twinned Class II icosahedral Goldberg-type nanodiamonds and star-shaped hexagonal diamond nanorods (classical Brenner potential) caused by breakdown of $sp^3$ bonds between (111) $sp^3$ carbon layers are presented in Figures 6a and 6b, respectively.

At ab initio and classical MD (AIMD) simulations levels of theory, graphitization of both Class I and Class II icosahedral Goldberg-type multiply-twinned nanodiamonds leads to formation of sets of enclosed Class I ($C_{20}$, $C_{80}$, $C_{180}$, etc.) and Class II ($C_{60}$, $C_{240}$, $C_{540}$, etc.) fullerenes and corresponding multiply-twinned closed-shell graphene polygons. Due to the symmetry of parent multiply-twinned diamond particles, all adjacent $sp^2$ shells follow graphite-type AB mutual staking sequence, which allows one to classify them as quasi-spherical closed-shell multiply-twinned graphite crystals. Classical MD simulations reveal 2.059 and 3.480 Å average distances between the adjacent shells for Class I and Class II Goldberg polyhedra, respectively (see Tables S1 and S2). The DFT optimization of the sets of embedded Class I and Class II fullerenes provides close intershell distances (2.030 and 3.497 Å for Classes I and II, respectively).

AIMD simulations of Class I ($C_{280}$) and II ($C_{300}$) Goldberg-type nanodiamonds up to 3000 K revealed direct $sp^3 \rightarrow sp^2$ structural transformation of Class II forming $C_{60}$ fullerene embedded inside $C_{240}$, whereas Class I nanodiamond oscillates keeping $sp^3$ skeleton intact. As the temperature increases up to 2000 K, both embedded fullerene structures collapse and become disordered states. The AIMD movies are presented in SM Section.

The experimental X-ray diffraction (XRD) patterns of the mirror-ball shaped μm-size carbon clusters in the Chelyabinsk meteorite detected the graphite (002) reflection that represents 3.35 Å interplanar spacing. Theoretical Raman spectrum of Class II 2 closed-shell graphite model ($C_{60}@C_{240}$) demonstrates the main Raman peaks at 1300 and 1500 cm$^{-1}$ (Figure SI7-12 of SM Section) which correspond well to experimentally observed D and G modes of $sp^2$ carbon vibrations [41]. Comparing theoretical calculations with experimental results, the interlayer distance of Class II is similar to the observed value with 3.9% and 4.4% mean absolute errors for

MD and DFT simulations, respectively (Supplementary Table S7-2). Theoretical Raman spectrum of 2 shells Class I closed-shell graphite model ($C_{20}@C_{80}$, 2.059 Å intershell distance, Supplementary Table S7-1) shows disordered multiple peaks in the whole energy region.

The central core (hexacyclooctadecane (-$C_{18}H_{12}$-) Figure SI7-1D) of hexagonal star-shaped multiply twinned diamond nanorod (Figure SI7-7) consists of 6 fused hexagonal rings and could be formed by fusion of $C_2$, $C_4$ and $C_6$ clusters in carbon-rich plasma [42, 43] during bolide flyby through Earth atmosphere. Following polyhexacyclooctadecane symmetry, multiply twinned elongated star-shaped hexagonal carbon dodecahedron is the natural product of $sp^3$ structural expansion of (-$C_{18}$-)$_n$ embryo [39]. Consequent graphitization of the multiply twinned star-shaped carbon dodecahedrons gives hybrid multiply twinned closed-shell graphite microcrystals of hexagonal cross section (Figure 6b, Figures SI7-8, SI7-11) which consist of embedded $sp^2$/$sp^3$ honeycombs in AB graphite staking sequence with 3.014 Å average intershell distances (Supplementary Table S7-3).

Joint consideration of the experimentally observed morphology of unique closed-shell quasi-spherical and elongated hexagonal symmetry graphite microcrystals with structural analysis and DFT and MD simulations provides the key insight of structure and formation mechanisms of closed-shell multiply-twinned graphite microcrystals. Symmetry, intershell distances and Raman spectra directly support $C_{60}$ fullerene as the main candidate for formation embryo of quasi-spherical graphite microcrystals through graphitization of icosahedral multiply-twinned nano- and microdiamonds. The high-temperature evolution of diamond microcrystals leads to a formation of quasi-spherical embedded $sp^2$ carbon shells which can be regarded as closed-shell multiply twinned graphenes or fullerenes of $C_{60}$ core (*n*, *n*) icosahedral symmetry and graphite AB mutual stacking. The elongated graphite closed-shell microcrystals of hexagonal symmetry seem to be the result of high-temperature evolution of hexagonal star-shaped multiply twinned diamond dodecahedrons with polyhexacyclooctadecane (-$C_{18}$-) core through formation of the embedded hybrid $sp^2$/$sp^3$ carbon honeycombs with AB mutual stacking. In both cases the intershell distances are very close to graphite interlayer distances.

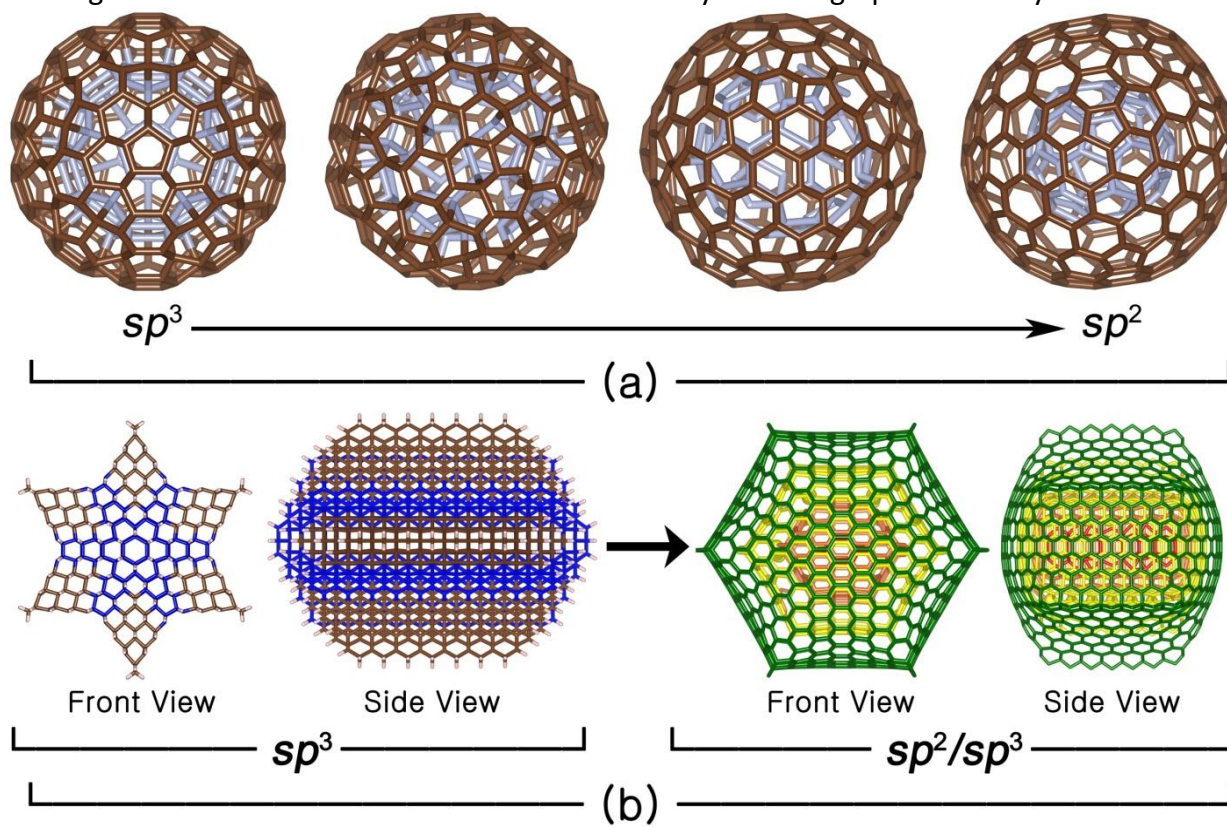

Figure 6. (a) Ab initio molecular dynamics simulation of $sp^3 \rightarrow sp^2$ structural transformation of icosahedral Class-II $C_{60}C_{240}$ nanodiamond to two enclosed $C_{60}@C_{240}$ fullerenes. (b) Top and side views of $sp^3 \rightarrow sp^2$ structural transformation of hexagonal star-shaped multiply-twinned diamond $C_{2016}$ nanorod with 3 $sp^3$ (111) layers around $(C_{18})_7$ hexacyclooctadecane core to 3 closed-shell embedded hybrid $sp^2/sp^3$ honeycombs with $sp^3$ $(C_{18})_7$ core in the center. Blue and brown colors represent boat-type (Lonsdaleite) and chair-type (diamond) configurations, respectively.

Unique structures found in the meteoritic dust that have not been observed before demonstrate the unlimited potential of nature to synthesize new materials. Mineral fibers, alongside spherules, can participate in the identification of meteorites that fell in the past and can point to this type of meteorites.

Pure carbon reduced from carbon dioxide as a result of thermal dissociation presumably crystallized on the surface of carbonaceous nanostructures (fullerenes and nanotubes) forming graphite stacks that repeat the spatial symmetry of the crystallization center. The conducted theoretical study shows that such structure is energetically stable and actually can form in the conditions observed during a bolide's flight. Such carbon crystals are unique and have not been observed before. According to our assessments, the concentration of these crystals along the trajectory of the Chelyabinsk superbolide's fall is about 10..100 pcs/m$^2$. Taking into account the spatial distribution of the gas-dust tail above Earth's Northern Hemisphere, such particles must have fallen out in great quantities in Russia, European and North American countries and can be found, in particular, in the snow of Alaskan and Kamchatkan mountains.

The analyzed carbon inclusions found in the rock sections of the fragments of Barringer's meteorite (lonsdaelite was first found in it) do not have the signs of crystallinity at micrometer and submicrometer dimensions. The supplementary materials present the image and elemental analysis of several fragments containing carbon inclusions.

It is necessary to note that any attempts to synthesize similar materials by the method of directed explosion and by comparing them with carbon structures from the known meteorites have failed so far. Detonation diamonds that were produced by the method of a directed explosion of graphite have not shown anything similar either. The supplementary materials present an image of such detonation diamond particles, their elemental microanalysis and data of X-ray structure analysis. Agglomerates, which were formed as a result of a directed explosion of graphite, contained diamonds of submicrometer dimensions. As follows from the presented data, detonation diamonds do not have similar morphology at the observed size scale of several dozen micrometers.

Consideration of both experimentally observed morphology and theoretical DFT and MD simulations of unique exotic quasi-spherical and elongated hexagonal symmetry graphite microcrystals provides clear insight of atomic structure and formation mechanisms of carbon microcrystals. High-temperature formation of multiply-twinned icosahedral Class-II diamonds with $C_{60}$ fullerene cores and star-shaped hexagonal dodecahedron diamonds (polyhexacyclooctadecane (-$C_{18}$-) core) with consequent $sp^3 \rightarrow sp^2$ structural evolution leads to formation of closed-shell icosahedral and elongated hexagonal symmetry graphite microcrystals with graphite-type AB mutual stacking of multiply-twinned closed-shell graphine shells. Symmetry analysis, intershell distances and theoretical Raman spectra of embedded $C_{60}@C_{240}$ and $C_{20}@C_{80}$ fullerenes directly support proposed structure and formation mechanism of unique exotic quasi-spherical and elongated hexagonal symmetry graphite microcrystals.

Microscopic analyses (conducted in 2014-2018) of the insoluble component of snow collected in the same areas as the samples of 2013 showed the absence of the materials described in this paper.

**Methods**

SEM and EDX were performed on JEOL 6510 LV electron microscope with EDX Oxford Instruments setup. As far as samples of collected dust were mainly consist of poor conductors or insulators they were coated with fine layer of platinum to prevent charging of the specimen. The thickness of the platinum layer does not exceed a few nanometers. Raman spectra and optical images were collected on Thermo Fisher Scientific DXR Raman Imaging Microscope, laser wavelength 532 nm, radiation power 1 mW. XRD patterns were collected on Oxford Xcalibur Single Crystal X-ray Diffractometer with graphite monochromated Mo Kα radiation (λ = 0.71073 Å) an inbuilt Sapphire CCD camera. All measurements were taken with long exposures (~700 s) with a turn of the particle by 1° after each exposure. XRD measurements of the detonation products were made using Rigaku Ultima IV diffractometer based on the Cu Kα radiation. Geometry optimization of the Class I and II Goldberg polyhedral was performed using both molecular dynamic simulation (MD) and density functional theory (DFT). Structures of star-shaped hexagonal sp3 dodecahedron and elongated hexagonal hybrid $sp^2$-$sp^3$ carbon honeycomb shells of elongated hexagonal closed-shell graphite microcrystal were optimized using Large-scale Atomic/Molecular Massively Parallel Simulator (LAMMPS) [44] molecular dynamics code based on Brenner potential [45]. The molecular structures of $sp^2$ Goldberg–type carbon polyhedra of Class I and II were built up to 10 shells applying Tersoff potential to prepare high symmetrical geometry. All combined shells were optimized together using molecular dynamics (MD) based on classical Brenner potential. For more accurate observation, three inner shells were calculated using Open source package for Material eXplorer (OpenMX) [46] DFT package. General gradient approximation (GGA) and Perdew, Burke, Ernzerhof (PBE) [47] with Grimme D3 Van deer Waals correction (DFT-D3) [48, 49] was employed to run electronic structure calculations. The energy cutoff was set to 75 Hartree within $10^{-10}$ Hartree convergence for self-consistent field (SCF) loop and optimized with the condition of $10^{-4}$ Hartree/Bohr optimization criterion and 300 K electronic temperature. The thermal trajectories of the $sp^3$ type Class I and II of 2 shells were simulated using ab initio molecular dynamics. The canonical (NVT) [50, 51] ensemble and Nosé-Hoover thermostat [50, 52, 53] were applied for MD simulations. The temperature was gradually increased from 300 to 3000 K within 25 ps total simulation time and 5 fs time interval. The Raman spectra of the Class I and II up to 2 shells were calculated using General Atomic and Molecular Electronic Structure System (GAMESS) [54, 55] code. The Austin model 1 (AM1) [56] parameterized semi-empirical approach was used to calculate Hessian matrix for 300 atoms model and Becke 3-parameter Lee, Yang, Parr (B3LYP) [57, 58] DFT approach with 3-21G [59] basis set was used to obtain derivative polarizability tensor.

## Acknowledgements

We thank all CSU students and employers who involved to the process of meteoritic dust collection. This work was partially supported by the RSF-Helmholtz project 18-42-06201 and S.T. acknowledges RFBR grant No. 16-07-00679, Act 211 Government of the Russian Federation, contract No. 02.A03.21.0011 and Ministry of Education and Science of the Russian Federation in the framework of Increase Competitiveness Program of MISiS. W.B., A.K. and P.A. acknowledge National Research Foundation of Republic of Korea for support under grant NRF-2017R1A2B4004440.

## Author contributions

S.T. discovered carbon particles, planned and supervised the whole project, wrote the paper. K.S., A.D.& O.G. provided scientific support and contributed to the discussions. G.S.& S.T. made optic investigations and all samples preparations. N.G. discovered mineral whiskers and organized with A.D. collection of meteoritic dust. W.D & T.F. performed XRD investigations and particles extraction. W.B. and A.K. performed DFT and MD simulations of multiply twinned carbon nano- and microcrystals, P.A. suggested structural models of multiply twinned carbon nano- and microcrystals and suggested the mechanisms of their formation. V.K.& S.T. made SEM and Raman measurements and analyze them. All the authors contributed to the revisions.

## Additional information

*Supplementary Information accompanies this paper at 62 pages.*

**Competing interests:** The authors declare no competing interests.